%% file: MAIN.tex
\documentclass[letterpaper, 10 pt, conference]{ieeeconf}  

\IEEEoverridecommandlockouts                              

\usepackage{graphicx}
\graphicspath{ {./Images/} }
\usepackage{graphics} 
\usepackage{epsfig} 
\usepackage{mathptmx} 
\usepackage{times} 
\usepackage{amsmath} 
\usepackage{amssymb}  
\usepackage{float}
\usepackage{array,booktabs,arydshln,xcolor}

\usepackage{geometry}
\usepackage{cite}
\usepackage[normalem]{ulem}
\useunder{\uline}{\ul}{}

\title{\LARGE \bf
An Industrial Applicable Approach towards Design Optimization of a Mechanism: a Coronaventilator Case Study.}

\author{Abdelmajid Ben Yahya$^{1,2,*}$, Nick Van Oosterwyck$^{1,2}$, Jan Herregodts$^{3}$, Stijn Herregodts$^{3}$, Simon Houwen$^{4,5}$, \\
Bart Vanwalleghem$^{4,5}$, Stijn Derammelaere$^{1,2}$
\thanks{$^{1}$Department of Electromechanics, CoSysLab, University of Antwerp, Antwerp, Belgium}%
\thanks{$^{2}$AnSyMo/CoSys, Flanders Make, the strategic research centre for the manufacturing industry, Belgium}%
\thanks{$^{3}$Department of Human Structure and Repair, Ghent University, Gent, Belgium}%
\thanks{$^{4}$Department of Electrical Energy, Metals, Mechanical Constructions and Systems, Ghent University Campus Kortrijk, Kortrijk, Belgium}%
\thanks{$^{5}$EEDT-MP, Flanders Make, the strategic research centre for the manufacturing industry, Belgium}%
\thanks{$^{*}$Corresponding author: abdelmajid.benyahya@uantwerpen.be}%
}

\begin{document}
\newgeometry{top=50pt,left=46pt,right=46pt,bottom=74pt}

\maketitle
\thispagestyle{empty}
\pagestyle{empty}

\begin{abstract}

\input{Sections/Abstract}
\end{abstract}

\section{Introduction}
\input{Sections/Introduction}

\section{CAD Motion Simulations}
\input{Sections/CAD_motion_simulation}

\section{Design Optimization}
\input{Sections/Optimization_approach}

\section{Results}
\input{Sections/Results}

\section{Conclusion}

\input{Sections/Conclusion}

\addtolength{\textheight}{-11cm}   





 \bibliographystyle{IEEEtran}
 \bibliography{library}

\end{document}

%% file: Sections/Abstract.tex
Design optimization of mechanisms is a promising research area as it results in more energy-efficient machines without compromising performance. However, machine builders do not actually use the potential described in the literature as these methods require too much theoretical analysis.

This paper introduces a convenient optimization workflow allowing wide industrial adoption, by using CAD models. The 3D multi-body software is used to perform motion simulations, from which the objective value samples can be extracted. These motion simulations determine the required torque for a certain combination of design parameters to fulfill the movement. Dedicated software can execute multiple motion simulations sequentially and interchange data between the different simulations, which automates the process of retrieving objective value samples. Therefore, without in-depth analytical design analysis, a machine designer can evaluate multiple designs at low cost. Moreover, by implementing an optimization algorithm, an optimal design can be found that meets the objective. In a case study of a coronaventilator mechanism with three design parameters (DP's), 39 CAD motion simulations allowed to reduce the RMS torque of the mechanism by 57.2\% in 42 minutes. 

%% file: Sections/Introduction.tex
\label{sec:Introduction}

\begin{figure}[b]
    \centering
    \includegraphics[width=1\columnwidth]{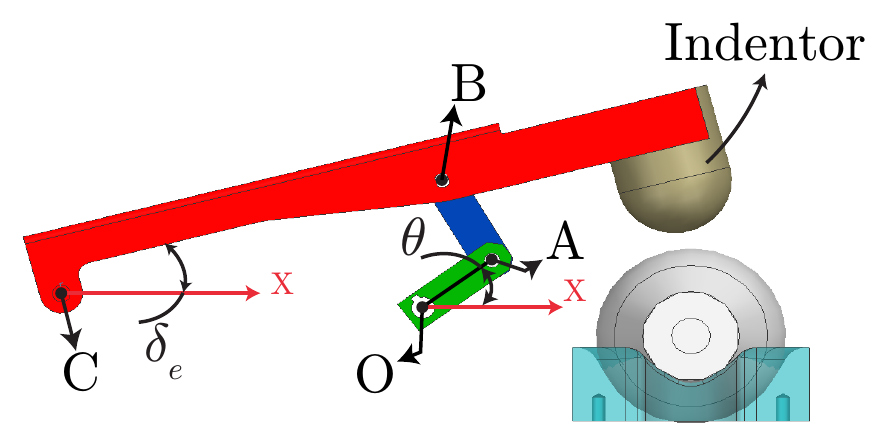}
    \caption{The considered design parameters $\vert OA \vert$, $\vert AB \vert$ and $\vert BC \vert$ of a coronaventilator, in the present paper.}
    \label{fig:Coronaventilator}
\end{figure}

\begin{figure*}[h]
    \centering
    \includegraphics[width=1.9 \columnwidth]{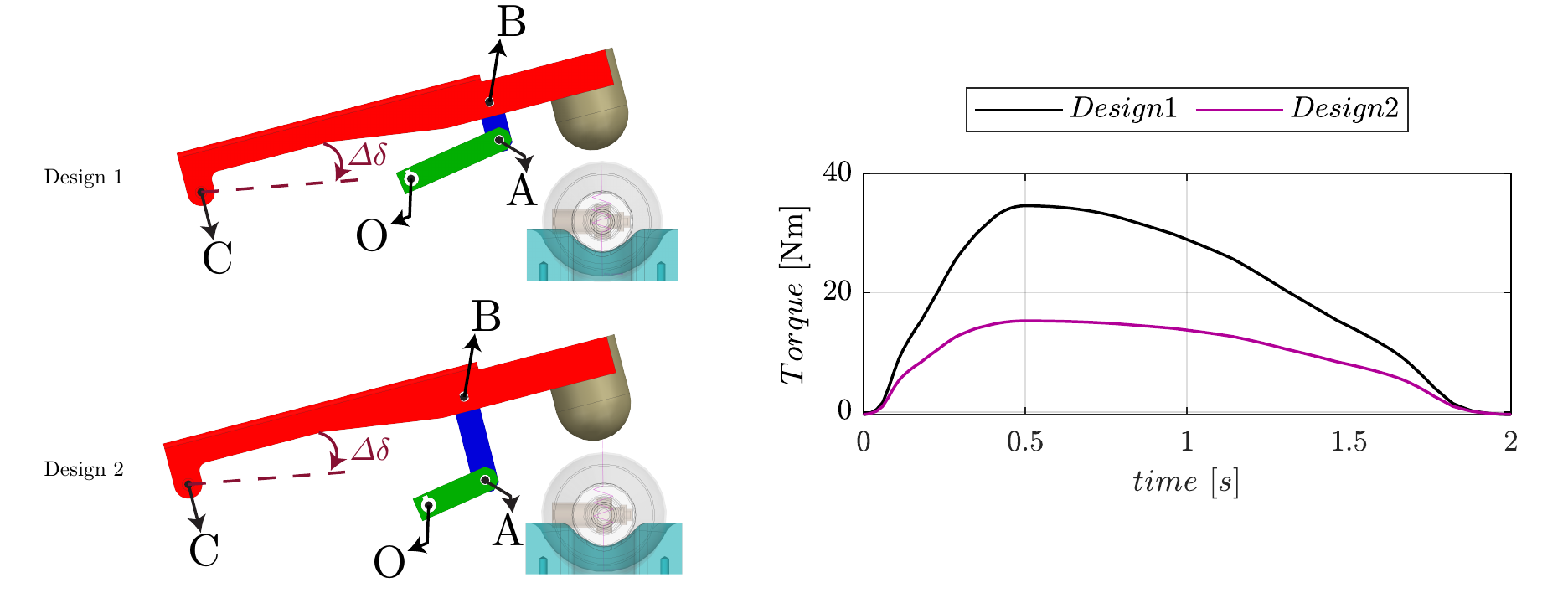}
    \caption{Defining specific lengths for the mechanism links influences the required speed and torque to move the end-effector (red beam) over a range of $\Delta\delta$, driven from point O.}
    \label{fig:optimconcept}
\end{figure*}

The energy consumption of industrial machinery is a topic of primary importance due to environmental and economic considerations \cite{Dornfeld2013}. The 45\% share that electric motors have in the global electric consumption \cite{Waide2011} supports the statement that any energy-saving method should be investigated thoroughly. The methodology proposed within this paper is applicable for all mechanisms with an imposed movement of the end-effector or tooltip. Many recent studies \cite{Carabin2017} pay attention to minimizing the energy dissipation in the electric motor to reduce the consumed electrical energy. Moreover, reducing the energy losses in the motor lowers the probability that the motor can be overheated \cite{VanOosterwyck2020}.
The link lengths in a mechanism can differ while fulfilling the same task, being the Point-To-Point (PTP) displacement of the end-effector. Therefore, within this case the geometry parameters $\vert OA \vert$, $\vert AB \vert$, and $\vert BC \vert$ of the coronaventilator depicted in Fig. \ref{fig:Coronaventilator} can be considered as design parameters to enhance the mechanism, while maintaining the imposed movement of the indentor. 
Design optimization of a PTP mechanism is one specific approach to reduce the energy consumption of electric machinery. As indicated in Fig. \ref{fig:optimconcept}, changing the geometry parameters ($\vert OA \vert$, $\vert AB \vert$ and $\vert BC \vert$) can result in a lower RMS torque (\(T_{RMS}\)). The literature states that minimizing the \(T_{RMS}\) corresponds with reducing the energy losses in the system \cite{Berselli2016}. Awareness about the influence of machine components geometry on energy consumption has recently attracted attention \cite{Carabin2017,Mashimo2015,Sheppard2019}. Mechanism models \cite{VanOosterwyck2020,Oosterwyck2019} replace the prototyping, allowing computational evaluation of multiple designs with limited cost.

A coronaventilator is used as a proof of concept within this study. This mechanism was constructed during the first wave of the covid-19 pandemic by a non-profit organization \cite{Herregodts2019}. Having continuous (24/7) electricity access is not obvious in low- and middle-income countries. Thus, having a mechanism that consumes a minimum of electric energy enabling the usage of batteries is highly relevant. Therefore, the objective of this study is to find the optimal design (being lengths $\vert OA \vert$, $\vert AB \vert$ and $\vert BC \vert$ in Fig. \ref{fig:Coronaventilator}) leading to a minimal $T_{RMS}$ for this PTP mechanism. As of now, the industry heavily relies on 3D multi-body software to design a mechanism. The method introduced in this paper uses these CAD models to sample the objective value through motion simulations. Furthermore, multiple motion simulations must be set up in a specific order and interchange information to have the correct objective value for a combination of design parameters $\vert OA \vert$, $\vert AB \vert$, and $\vert BC \vert$. The automation of the process in which multiple simulations are run sequentially is performed by dedicated software \cite{HEEDS}. State-of-the-art design optimization methods derive the torque for a mechanism's movement analytically \cite{El-Kribi2013}, which is mechanism specific and less convenient. However, the methodology described in this paper only needs CAD software, which ensures broad industrial applicability.

\subsection{Related Work}
\label{sec:Related work}
In the literature \cite{Affi2007}, the minimization of the driving torque is done by establishing dynamic equations of the system to predict the dynamics. Moreover, \cite{El-Kribi2013,Affi2007,Rayner2009} do not define the feasible search domain nor include it in searching for the optimal result. Indicating the feasibility of a certain design is important as defects, giving infeasible designs, \cite{Hernandez2021} can occur in the synthesis of a mechanism. The optimization algorithms of \cite{El-Kribi2013,Affi2007,Gogate2012} assure that the objective function has converged towards a minimum, yet it is generally not guaranteed that the designed linkage will be feasible. Therefore, the necessary analysis should be added so that the optimal solution can fulfill the movement without issues.

Developing a reciprocal mechanism that follows the desired end-effector trajectory is a classical design problem that researchers extensively explore \cite{Cabrera2002}. However, all methods in the literature \cite{El-Kribi2013,Affi2007,Rayner2009,Hernandez2021,Gogate2012,Cabrera2002} use dynamic equations, which are case specific and inconvenient for industrial applicability. Therefore, this paper aims to describe a workflow on optimizing a mechanism through the usage of motion simulations and dedicated software that automates this methodology.

\subsection{Method}
\label{sec:Method}

This paper describes the methodology one should apply to optimize a mechanism through CAD-based motion simulations. Mechanical design of systems is mainly done in Computer-Aided Design (CAD) software. These CAD models include all required information (i.e., volume, mass, friction, damping, joints,...) to model the dynamics of a mechanism. This information is necessary to calculate the required torque of the mechanism through motion simulations. By driving the mechanism with the motion profile $\theta(t)$ at point O (Fig. \ref{fig:Coronaventilator}), being the axis driven by a motor, the user can extract the necessary torque from the software (as in Fig. \ref{fig:optimconcept}) to fulfill the prescribed movement $\delta(t)$ of the end-effector. The objective value is the \(T_{RMS}\), necessary to drive the mechanism fulfilling an imposed PTP motion ($\delta(t)$). 

Hence, by calculating the \(T_{RMS}\) based on CAD simulations as elucidated in Section \ref{sec:CAD motion simulation}, the objective value for a particular design (i.e., specific values for the three design parameters $\vert OA \vert$, $\vert AB \vert$, and $\vert BC \vert$) is obtained. Be aware that changing the geometric parameters ($\vert OA \vert$, $\vert AB \vert$, and $\vert BC \vert$) influences the start- and end angle of the motor $\theta$ as the imposed end-effector movement $\delta(t)$ may not change. Therefore, a kinematic transformation is necessary to derive the motor profile $\theta(t)$ for a specific movement of the end-effector $\delta(t)$. However, performing a kinematic transformation for a design parameter combination ($\vert OA \vert$, $\vert AB \vert$, and $\vert BC \vert$) that results in an infeasible design is irrelevant. Thus, all designs have to pass a feasibility check before any calculations are performed on the design. After obtaining the motion profile of the motor ($\theta(t)$), for a specific feasible design parameter combination, it can be used as a driver of the mechanism in simulation to perform a dynamic analysis and derive the required torque profile. The whole process described above is a sequence of motion simulations that is automated in order to obtain the objective value for different feasible design parameter combinations ($\vert OA \vert$, $\vert AB \vert$ and $\vert BC \vert$). As the objective value $T_{RMS}$ for a certain design of the mechanism can be derived, an optimization algorithm can be used, as discussed in section \ref{sec:Optimization approach}. This algorithm is necessary to minimize the \(T_{RMS}\) and thus optimize the mechanism. After optimization, optimal combination of the design parameters $\vert OA \vert$, $\vert AB \vert$, and $\vert BC \vert$ is obtained that has a minimal $T_{RMS}$, and therefore consumes a minimal amount of energy, as shown by results in section \ref{sec:Results}.

\begin{figure*}[t]
    \centering
    \includegraphics[width=1.8\columnwidth]{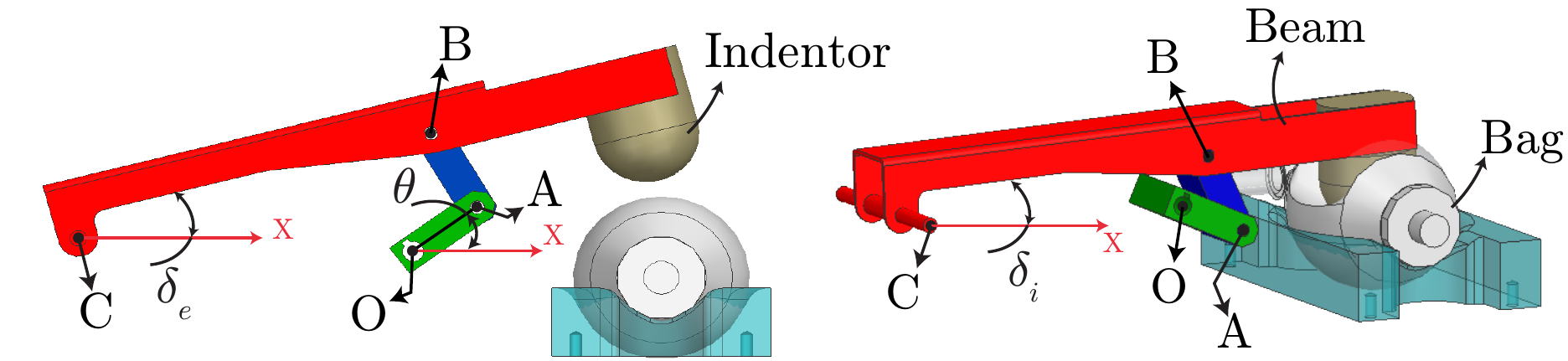}
    \caption{The end-effector (indentor) requires a movement from $\delta_i$ to $\delta_e$, which is performed by moving $\theta$ over a design-specific angle.}
    \label{fig:coronaventilator movement}
\end{figure*}

\begin{figure*}[t]
    \centering
    \includegraphics[width=1.8\columnwidth]{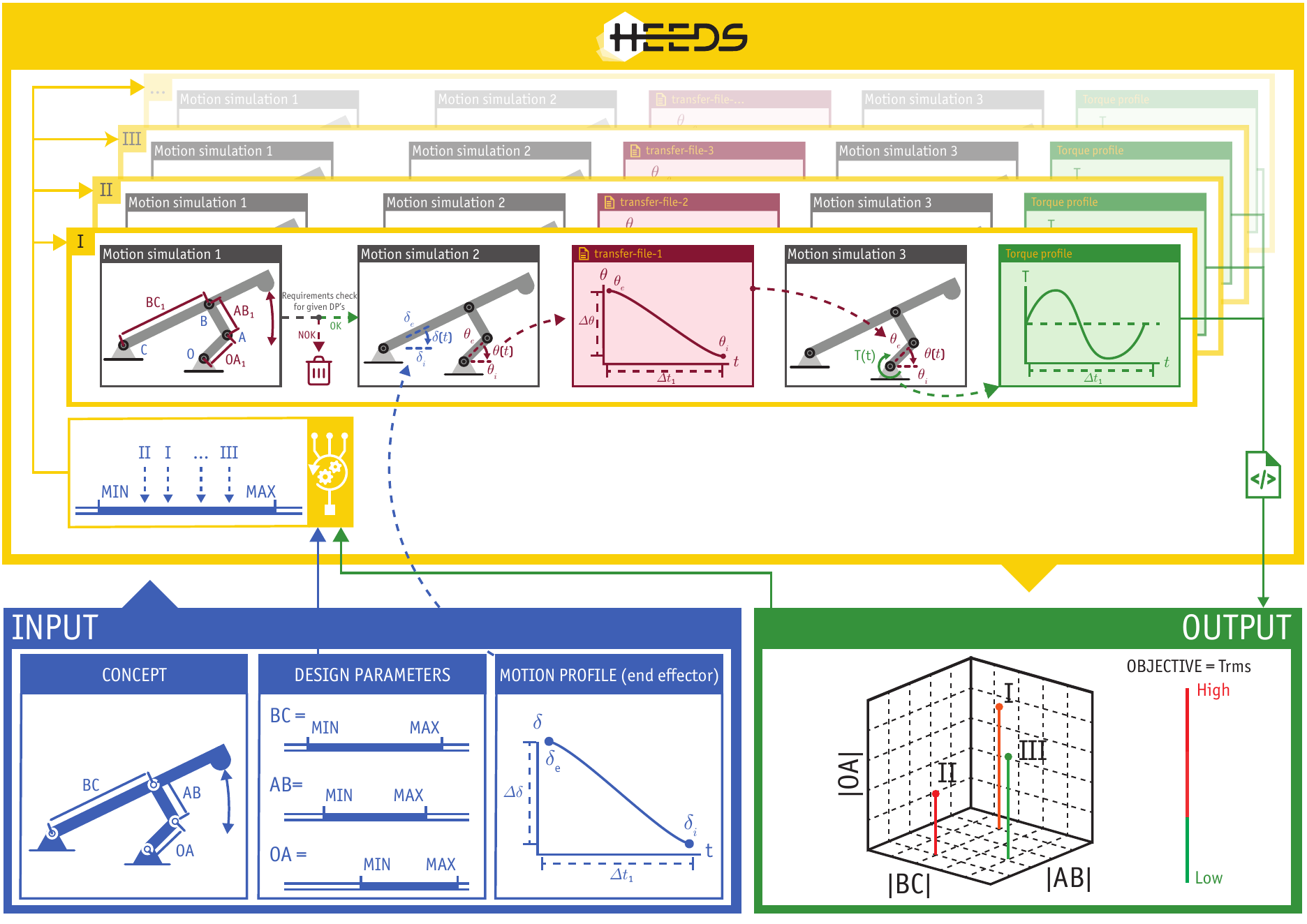}
    \caption{The workflow for automated extraction of the necessary driving torque for different feasible mechanism designs. One can optimize the mechanism using an algorithm that chooses the design parameter combinations of what  the objective value must be determined.}
    \label{fig:workflow}
\end{figure*}   

%% file: Sections/CAD_motion_simulation.tex
\label{sec:CAD motion simulation}

The validation case is clarified to make all the following more tangible. This mechanism, shown in Fig. \ref{fig:coronaventilator movement}, can ventilate a patient by pressing the indentor into the bag, which causes airflow towards the patient. The movement of the end-effector (indentor) from a starting angle $\delta_i$ towards an end angle $\delta_e$ is caused by moving point O over $\theta(t)$. In this paper, the machine designer only defined an end-effector (indentor) motion profile $\delta(t)$, resulting in a reciprocal movement between the positions $\delta_i$ and $\delta_e$.

Fig. \ref{fig:coronaventilator movement} presents the CAD model of the coronaventilator and illustrates that the red beam, connected with the indentor (i.e., the end-effector), moves by rotating input link OA around point O. This is the point where an electric motor drives the mechanism. The red beam has two predefined angles: an angle $\delta_e$ that holds the mechanism in a position where the indentor touches the bag and an angle $\delta_i$ that corresponds to a position in which the air is compressed out of the bag. Within these CAD models, the design parameters $\vert OA \vert$, $\vert AB \vert$, and $\vert BC \vert$ of the coronaventilator can be parameterized to simulate different designs with different corresponding torque profiles, as shown in Fig. \ref{fig:optimconcept}.

A CAD motion simulation \cite{SiemensNX} can determine the necessary torque to drive the mechanism at point O only if the required position profile $\theta(t)$, at that point O, is known. However, the user solely defines the required motion profile of the end-effector, in this case, $\delta(t)$. Thus, a machine designer should determine the specific motor angles $\theta(t)$ to move the end-effector through the imposed motion profile $\delta(t)$. It should be noted that the kinematic transformation from $\delta(t)$ to $\theta(t)$ depends on the chosen design parameter combination $\vert OA \vert$, $\vert AB \vert$, and $\vert BC \vert$. Moreover, each evaluated design must be feasible  to extract a representative objective value. Therefore, each selected design parameter combination ($\vert OA \vert$, $\vert AB \vert$ and $\vert BC \vert$) is analyzed by a sequence of three motion simulations that are executed automatically, as indicated in Fig. \ref{fig:workflow}.

\subsection{Motion Simulation 1: Feasibility Check}
\begin{figure}[b]
    \centering
    \includegraphics[width=1.0\columnwidth]{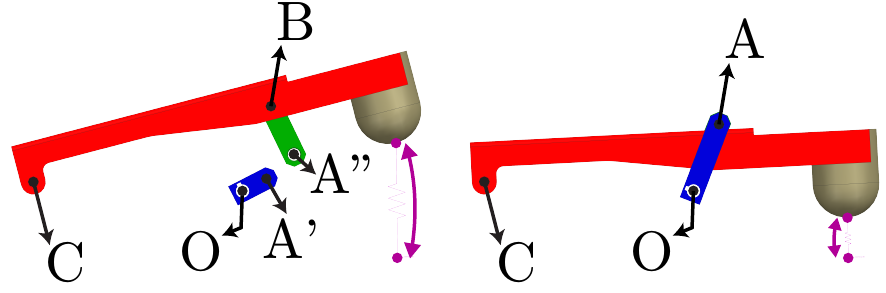}
    \caption{On the left, a coronaventilator design that cannot be assembled in the highest desired position of the complete range of motion. Another combination of design parameters results in a mechanism that cannot complete the desired range of motion, on the right.}
    \label{fig:unfeasible}
\end{figure}

As a first step within the series of motion simulations, each combination of design parameters ($\vert OA \vert$, $\vert AB \vert$ and $\vert BC \vert$) should be checked on feasibility. As depicted in Fig. \ref{fig:unfeasible} (left), a design parameter combination can result in an infeasible mechanism in which the link OA' cannot be connected with the link A"B at the highest position of its range of motion. Additionally, Fig. \ref{fig:unfeasible} (right) shows an infeasible design wherein the end-effector cannot reach the lowest desired position.  Thus, a first simulation is required to check the assemblability of each new mechanism over the required range of motion. 

\begin{figure}[t]
    \centering
    \includegraphics[width=1.0\columnwidth]{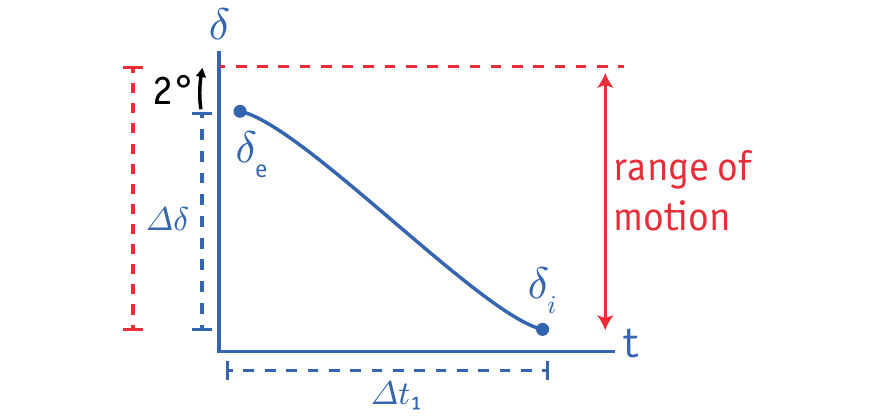}
    \caption{The machine designer requires a difference between the range of motion and the movement $\delta(t)$ to improve the ease of use.}
    \label{fig:RoM}
\end{figure}

The first motion simulation drives the CAD model from point C, the end-effector see Fig. \ref{fig:unfeasible}, over the desired range of motion. The output of this simulation is either false or true, which means that the simulated design is respectively infeasible or feasible. Infeasible design parameter combinations are neglected and neither used in the following simulations, nor by an algorithm to choose a better design, as shown in Fig. \ref{fig:workflow}. A specific range of motion can be a machine designer requirement for the machine, as it is in this case. The indentor has to move further up, so there is enough clearance to remove or place the bag. As shown in Fig. \ref{fig:RoM}, the range of motion is 2 degrees bigger than the actual $\delta(t)$ movement. However, when the range of motion is not explicitly desired the first simulation is removed, as the second simulation can give the same outcome. 

\subsection{Motion Simulation 2: Kinematic Transformation}
The feasible design is provided to the second motion simulation. As indicated in Fig. \ref{fig:workflow}, each design needs the complete motor's motion profile $\theta(t)$, to move the end-effector according to the imposed motion profile $\delta(t)$. This step is crucial as each design parameter combination requires another motion profile $\theta(t)$ at the motor (point O) to preserve the same end-effector movement. It is possible to calculate the motion profile $\theta(t)$ analytically. However, deriving the kinematic equations is complex and  can only be used for a particular mechanism. For this reason, CAD motion simulation is used to perform complex calculations instead of manual calculation by machine builders. The CAD model is driven from the end-effector (point C) with the desired motion profile $\delta(t)$. Because of the kinematic transformation that the CAD software executes on the mechanism, the user can extract the corresponding motor profile $\theta(t)$ for a specific design (Fig. \ref{fig:workflow}). Subsequently, the motor profile is used in the next and last motion simulation.

\subsection{Motion Simulation 3: Dynamic Analysis}
As shown in Fig. \ref{fig:workflow}, the last simulation determines the required driving torque of a specific design. The design that just passed through the previous simulation and the design-specific motor motion profile is provided to this third simulation. The CAD model drives the mechanism, as in real-life, from point O. As a consequence of the dynamic analysis the CAD software performs during such a simulation, one can extract the required torque the motor should provide to drive a specific design of the mechanism as desired. Based on the design-specific torque profile, the $T_{RMS}$ objective value for each design can be calculated. 

The complete sequential process with the three motion simulations makes it possible to extract the objective value for every feasible design. However, this workflow has to run automated to optimize a mechanism through algorithms. Therefore, the present paper utilizes HEEDS MDO \cite{HEEDS} as commercial software to automate this workflow and has most common optimizers integrated. As explained in section \ref{sec:Optimization approach}, algorithms are used to optimize the mechanism.

%% file: Sections/Optimization_approach.tex
\label{sec:Optimization approach}

\begin{figure}[t]
    \centering
    \includegraphics[width=1.0\columnwidth]{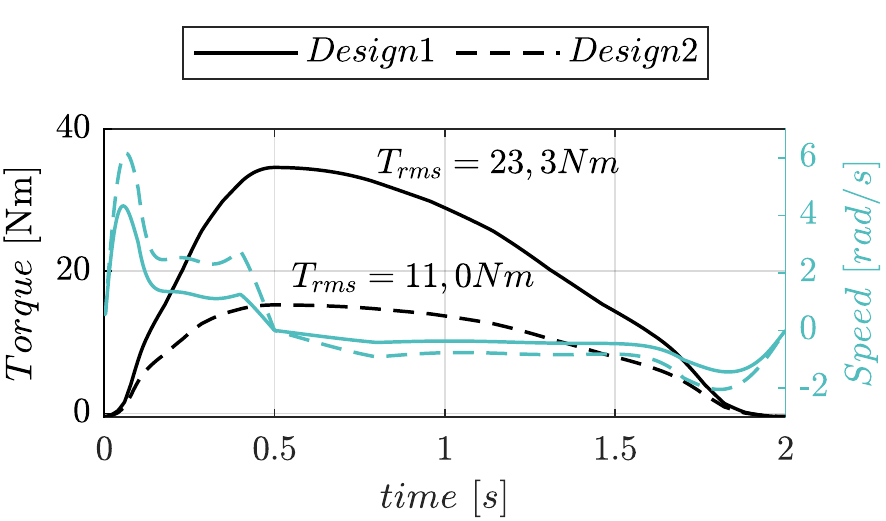}
    \caption{A design with a lower $T_{RMS}$ demands a higher maximal speed from the motor.}
    \label{fig:TRMS_Speed}
\end{figure}

An optimization algorithm uses the obtained objective value for a specific design of the mechanism to create a new design parameter combination, which possibly improves our objective value and gets closer to the optimal design with a minimal $T_{RMS}$. Yet, notice that a design with a lower \(T_{RMS}\) requires a higher maximal speed of the motor, as shown in Fig. \ref{fig:TRMS_Speed}. However, the increased motor speed stays within a realistic range, thus giving it no further focus. 

The optimization of the mechanism is an iterative method. Therefore, automating the sequence of motion simulations is a crucial step. The two most commonly used algorithms in design optimization \cite{Hernandez2021} are the Sequential Quadratic Programming (SQP), and the Genetic Algorithm (GA). Therefore, a comparative study is conducted between these two algorithms. The different algorithms that search for the optimal design parameter combination $\vert OA \vert$, $\vert AB \vert$ and $\vert BC \vert$ are SQP as a gradiënt-based method and GA as a heuristic optimizer.

\begin{itemize}

    \item \textbf{Genetic algorithm:} A GA is an adaptive heuristic search method based on the evolution of genetics. The algorithm starts with a population with a determined amount of random designs. All the designs in a generation are evaluated and get a fitness value, on which the next generation is created through evolution principles such as crossover, mutation, and selection. This evolutionary algorithm iteratively evolves towards improved objective values \cite{Bodenhofer2014,Kumar2020}. 

    \item \textbf{Sequential quadratic programming:} The basic idea of sequential quadratic programming is establishing an iterative procedure where the search of the optimal solution is led by the gradient of the quadratic model on the objective function in every design parameter combination. For each design, the objective function decreases fastest if the following design goes in the direction of the negative gradient of the objective function. The literature \cite{Boggs2000} describes this algorithm as one of the most successful in solving nonlinear constrained optimization problems. 
    
\end{itemize}

Thus, adopting the algorithms above on our mechanism will drive the process towards an optimal design parameter combination ($\vert OA \vert$, $\vert AB \vert$, and $\vert BC \vert$) for the coronaventialtor. The number of evaluations the algorithm performs must be limited as the sequence of motion simulations can be very time-consuming. However, taking the number of evaluations too low can lead to a poor result of the algorithm. Therefore, the algorithm can only stop when the objective value $T_{RMS}$ converged to a minimum.

%% file: Sections/Results.tex
\label{sec:Results}

\begin{figure}[t]
    \centering
    \includegraphics[width=1\columnwidth]{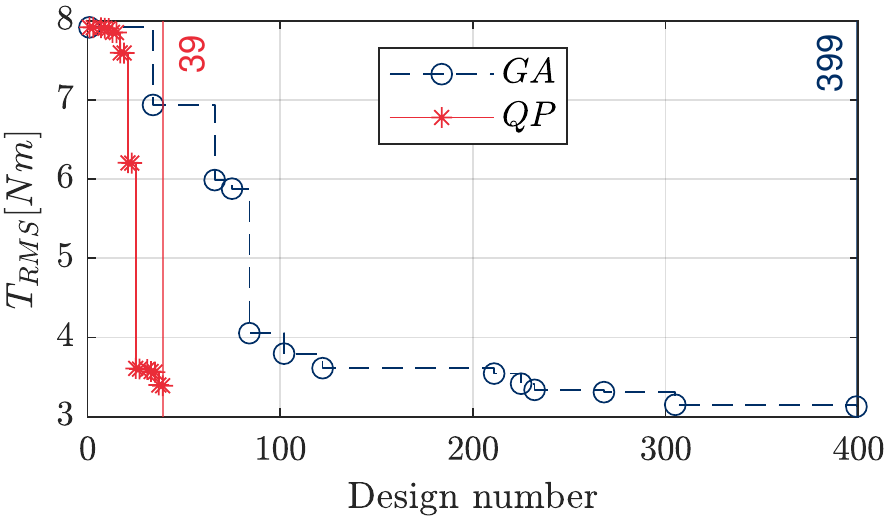}
    \caption{Both algorithms require a different amount of design evaluations to reach a minimal $T_{RMS}$ value. Moreover, the optimal objective value slightly differs for both algorithms.}
    \label{fig:ConvergencePlot}
\end{figure}

The method described in sec. \ref{sec:CAD motion simulation} is employed on the coronaventilator, which optimizes the mechanism by using an algorithm that searches for a new design parameter combination ($\vert OA \vert$, $\vert AB \vert$ and $\vert BC \vert$), as described in sec. \ref{sec:Optimization approach}. The first algorithm used in the methodology is the Genetic Algorithm, which found an optimal solution after 399 design evaluations, as indicated in Fig. \ref{fig:ConvergencePlot}. Within these evaluations, the algorithm chose some feasible and other infeasible designs. Each feasible design requires, on average, 1 minute and 5 seconds, while an infeasible design only takes 21 seconds. The time difference is a consequence of the methodology in which infeasible designs are detected in the first simulation and not passed on to the following simulation, as explained in sec. \ref{sec:CAD motion simulation}. The genetic algorithm evaluates 272 feasible and 127 infeasible designs, giving a total of 399 designs (see Table \ref{tab:SOTA_designoptimization}), to search for the optimal design requiring a calculation time of 5 hours and 40 minutes.

By contrast, the mechanism's optimal design through a gradient-based algorithm is found through a reasonable number of 39 feasible design evaluations, requiring 42 minutes of simulation time. It can be noticed that all designs chosen by the SQP algorithm are feasible designs, which is a consequence of our objective and the search method of the algorithm. The algorithm starts in a feasible design parameter combination ($\vert OA \vert$, $\vert AB \vert$ and $\vert BC \vert$) and chooses the next design, with a small increment, in the direction of the steepest negative gradient of the objective function.

\begin{table}[h]
\centering
\setlength\extrarowheight{5pt}
\caption{Saving potential achieved by design optimization with state-of-the-art optimization algorithms.}
\label{tab:SOTA_designoptimization}

\resizebox{0.48\textwidth}{!}{%
\begin{tabular}{ccccccccc}
Design   & \begin{tabular}[c]{@{}c@{}}$\vert OA \vert$\\ {[}mm{]}\end{tabular} & \begin{tabular}[c]{@{}c@{}}$\vert AB \vert$\\ {[}mm{]}\end{tabular} & \begin{tabular}[c]{@{}c@{}}$\vert BC \vert$\\ {[}mm{]}\end{tabular} & \begin{tabular}[c]{@{}c@{}}$T_{rms}$\\ {[}Nm{]}\end{tabular} & \begin{tabular}[c]{@{}c@{}}$T_{max}$\\ {[}Nm{]}\end{tabular} & \begin{tabular}[c]{@{}c@{}}$T_{rms}$\\ savings\\ {[}\%{]}\end{tabular} & \begin{tabular}[c]{@{}c@{}}$T_{max}$\\ savings\\ {[}\%{]}\end{tabular} & \begin{tabular}[c]{@{}c@{}}$Number$\\ $of$\\ $evaluations$\end{tabular} \\ \specialrule{0.5pt}{0pt}{0pt}
Original & 53                                                                  & 65                                                                  & 282                                                                 & 7.91                                                         & 13.26                                                        & -                                                                      & -                                                                      & -                                                                   \\
GA       & 82.68                                                               & 141.25                                                              & 281.8                                                               & 3.13                                                         & 5.43                                                         & 60.5                                                                   & 59.1                                                                   & 399                                                                 \\
SQP       & 30                                                                  & 76.22                                                               & 271.75                                                              & 3.39                                                         & 5.16                                                         & 57.2                                                                   & 61.1                                                                   & 39                                                                 
\end{tabular}%
}
\end{table}

However, both algorithms lead to different optimal design suggestions, which are better than the original design suggested by a machine builder. The optimal objective value found with the SQP algorithm is slightly higher than the solution obtained with GA, as shown in Table \ref{tab:SOTA_designoptimization}. Yet, the contrast between the two algorithms is noticeable in the required number of design evaluations to find an optimal design. In summary, Table \ref{tab:SOTA_designoptimization} shows that GA was able to reduce the $T_{RMS}$ by 60.5\%, while SQP diminished the objective value by 57.2\%. The optimal design found through a gradient-based method is quick, yet the obtained $T_{RMS}$ value strongly depend on the combination of $\vert OA \vert$, $\vert AB \vert$, and $\vert BC \vert$, in which the algorithm starts searching (i.e. starting point). This reveals that by using the SQP algorithm a risk is taken of having an sub-optimal design, which is a local optimum. In addition, both algorithms were able to lower the maximal torque the motor should deliver during the mechanism's movement, which means that the mechanism can operate with a smaller, and thus cheaper motor.

%% file: Sections/Conclusion.tex
This study proposes a convenient and broad industrial applicable design optimization approach, in which CAD models are used as a basis. The workflow requires multiple motion simulations of the CAD model to extract the necessary torque to drive the designed mechanism. The methodology described in this paper does not demand any theoretical analysis of the mechanism, making it general and applicable for any PTP mechanism in the industry. The first motion simulation checks the feasibility of the proposed design parameter combination ($\vert OA \vert$, $\vert AB \vert$, and $\vert BC \vert$). Then only the feasible designs are passed to the second simulation, which performs the kinematic transformation to derive the design-specific motor profile for an imposed end-effector movement. At last, the derived motor profile is used in the last motion simulation to perform dynamic analysis and extract the necessary torque to drive the mechanism with the chosen design. The obtained objective value ($T_{RMS}$) from the torque profile is used by the optimization algorithm to choose an improved design, which is evaluated by the sequence of motion simulations. 

The results clearly show that the proposed method outperforms the arbitrary designs chosen by the machine builder and reveals an energy-saving potential up to 60.5\%. Moreover, the choice of an algorithm significantly influences the number of designs evaluated to find an optimal design for the mechanism. The gradient-based algorithm only needed 39 design evaluations, which benefits the method's applicability.